\documentclass[10pt]{article}
\usepackage{fullpage,enumitem,amsmath,amssymb,graphicx,amsthm,listings}
\usepackage{blkarray, mathtools, tikz, lipsum, bbm}
\usepackage[ruled]{algorithm2e}
\usepackage{hyperref}
\usepackage{subcaption}
\usepackage{soul}
\hypersetup{
     colorlinks   = true,
     linkcolor    = blue,
     citecolor    = red
}
\usepackage[
backend=biber,
style=ieee,mcite
]{biblatex}

\SetKwComment{Comment}{\# }{ }
\newtheorem{definition}{Definition}
\newcommand{\vecalt}[1]{\boldsymbol{#1}}

\newcommand{\abs}[1]{\vert #1 \vert}

\providecommand{\vlambda}{\ensuremath{\vecalt{\lambda}}}

\begin{document}

\begin{center}
{\Large  Evaluating the Impact of Multiple DER Aggregators on Wholesale Energy \\[2pt]
Markets: A Hybrid Mean Field Approach}\\[5pt]

Jun He$^{*}$ and Andrew L. Liu$^\dagger$ \\
$^*$ School of Industrial Engineering, Purdue University, West Lafayette, IN 47906. Email:
he184@purdue.edu \\
$^{\dagger}$ School of Industrial Engineering, Purdue University, West Lafayette, IN 47906. Email: andrewliu@purdue.edu \\

\end{center}

\section*{Summary}
The integration of distributed energy resources (DERs) into wholesale energy markets can greatly enhance grid flexibility, improve market efficiency, and contribute to a more sustainable energy future. As DERs -- such as solar PV panels and energy storage -- proliferate, effective mechanisms are needed to ensure that small prosumers can participate meaningfully in these markets. To address this, the Federal Energy Regulatory Commission (FERC) in the U.S. issued Order 2222 in 2020, to enable DERs to aggregate and participate in wholesale markets alongside traditional resources. However, detailed mechanisms are still under development, and there remains uncertainty about which approaches are most effective for ensuring small-scale participation.

In response to this challenge, we study a wholesale market model featuring multiple DER aggregators, each controlling a portfolio of DER resources and bidding into the market on behalf of the DER asset owners. The key of our approach lies in recognizing the repeated nature of market interactions the ability of participants to learn and adapt over time.  Specifically, Aggregators repeatedly interact with each other and with other suppliers in the wholesale market, collectively shaping wholesale electricity prices (aka the locational marginal prices (LMPs)). We model this multi-agent interaction using a mean-field game (MFG), which uses market information -- reflecting the average behavior of market participants -- to enable each aggregator to predict long-term LMP trends and make informed decisions. Unlike traditional game models, where the strategies of all individual agents must be considered, the mean-field approach focuses on the collective impact, simplifying the decision-making process.

For each aggregator, because they control the DERs within their portfolio under certain contract structures, we employ a mean-field control (MFC) approach (as opposed to a MFG) to learn an optimal policy that maximizes the total rewards of the DERs under their management. This approach accounts for various uncertainties, such as solar output variability, demand fluctuations, and LMP dynamics, allowing aggregators to learn how to better manage their resources and optimize their market participation strategies.



The innovation of our work lies in the combination of a hybrid MFG and MFC approach, which allows for the scalable modeling of multiple DER aggregators interacting within a wholesale market. This hybrid approach effectively captures both the decentralized decision-making of individual aggregators and their collective impact on market dynamics. We also propose a reinforcement learning (RL)-based method to help each agent learn optimal strategies within the MFG framework, enhancing their ability to adapt to market conditions and uncertainties. Unlike descriptive models that merely outline market participant behavior, our model is prescriptive, designed for control automation. With the installation of grid edge devices running our RL algorithms, market participants can automatically achieve the market outcomes described in this paper.

We validate our approach using an electricity network model based on Oahu Island, incorporating realistic data on solar panel output profile, penetration levels, and generator characteristics. Numerical simulations show that LMPs quickly reach a steady state in the hybrid mean-field approach. Furthermore, our results demonstrate that the combination of energy storage and mean-field learning significantly reduces price volatility compared to scenarios without storage. 

\textbf{Keywords:} Mean-field equilibrium; multi-agent reinforcement learning; distributed energy resource; wholesale energy market; locational marginal price; energy storage, solar PV, aggregators.

\section{Introduction}

The growing adoption of distributed energy resources (DERs), such as solar photovoltaic (PV) panels and energy storage systems, will lead to significant changes in the dynamics of electricity markets. Once considered peripheral, these resources are now central to discussions about enhancing grid flexibility, improving market efficiency, and promoting sustainability. The FERC Order 2222 \mcite{fercset, *ferc2222, *ferc2222-A, *ferc2222-B} issued in 2020 recognizes the potential of DERs to improve grid reliability and economic efficiency and encourages the participation of DERs in wholesale energy markets, allowing them to compete alongside traditional power resources. However, it also introduces significant challenges, particularly in ensuring meaningful participation for small-scale prosumers -- those who both produce and consume electricity.

While the FERC order lays the groundwork for DER-integration into wholesale markets, the detailed mechanisms are still under development, with different RTOs/ISOs proposing different approaches. There remains uncertainty about which approaches are most effective for integrating these resources into market operations. Current research predominantly focuses on how a single aggregator manages its portfolio of DER assets, often assuming that the whole energy prices (aka the locational marginal prices (LMPs)) are exogenously given and remain unaffected by the aggregator's actions. These models typically optimize the aggregator's strategy to maximize rewards based on these fixed LMPs, without considering the feedback loop between the aggregator’s decisions and market prices. 

In contrast, our work addresses a more complex setting where multiple aggregators operate within the same market, each capable of influencing LMPs endogenously. In this setting, the aggregators are engaged in a non-cooperative game, which we model as a mean field game (MFG). 
An MFG is a game-theoretic framework used to study the behavior of a large number of interacting agents whose decisions are influenced by the average behavior of the group. In an MFG, each agent, like an aggregator, optimizes its own strategy based on both its individual objectives and the `mean field,' which represents the average effect or state of all agents in the market. This allows each agent to account for how their actions, combined with those of many others, impact key market outcomes like prices. 

Another innovation in our approach is that within each aggregator, we employ a mean field control (MFC) approach, as opposed to a centralized optimization, to manage the portfolio of DERs. 
MFC focuses on finding a policy for a representative agent (the mean field) that, when applied across all agents, optimizes the overall collective outcome, rather than solving a complex optimization problem for each individual DER owner. This approach reflects a cooperative framework (in contrast to the non-cooperative interactions among different aggregators in an MFG), where the DER assets are treated as if they are working together to maximize their collective rewards. The MFC method is more scalable than centralized optimization and is well-suited for incorporating reinforcement learning (RL) techniques. This allows the aggregator to learn and adapt to uncertainties such as solar output variability, demand fluctuations, and LMP dynamics, ultimately developing optimal bidding strategies over time. 

By combining MFC for managing internal operations within each aggregator and MFG for modeling interactions among different aggregators, our model offers a comprehensive perspective on how multiple aggregators, each with their own portfolio of DERs, can strategically interact and impact market outcomes in a decentralized manner. Integrating these advanced modeling techniques with RL algorithms, our research presents a novel and scalable approach that is conducive to control automation. The RL algorithms designed for MFC can be embedded into smart controllers at DER owner sites, enabling efficient integration of DERs under the management of aggregators. This approach enhances market efficiency and stability, especially as DER adoption grows, by enabling aggregators to dynamically learn and adjust their strategies, taking into account both market interactions and inherent uncertainties.

While RL algorithms have been extensively studied and applied in various domains,  their traditional applications often focus on single-agent scenarios or environments with fully observable states. Our work, however, deals with multi-agent reinforcement learning (MARL), which introduces additional challenges, particularly in scalability and coordination among agents. The majority of existing MARL algorithms are designed for cooperative games, with only a few exceptions addressing non-cooperative scenarios. Regardless, the MARL algorithms often struggle with scalability due to the need for each agent to consider the state and actions of all other agents in the system. Both MFC and MFG offer specific ways to make MARL scalable, as agents update their policies based on the mean field -- a representation of the average effect of all agents—rather than relying on the detailed state information of every individual agent.
In this work, we also contribute a MARL algorithm tailored for our hybrid MFC-MFG approach, drawing on recent advancements in RL algorithms for MFC and MFG \cite{guo2019learning, mondal2023mean, gu2024mean} and proposing a two-phase learning approach.


The rest of the paper is organized as follows: Section \ref{sec:Wholesale} introduces the wholesale energy market model and explains the calculation of LMPs. In Section \ref{sec:Agg}, we outline the decision-making problem faced by aggregators. Section \ref{sec:alg} details our proposed two-phase learning algorithm. In Section \ref{sec:numerical}, we validate our approach using a case study of an electric network model based on Oahu Island. Finally, Section \ref{sec:Conclusion} concludes with a summary of our findings and potential directions for future research.

\section{Wholesale Market Model and Locational Marginal Prices}\label{sec:Wholesale}
In this section, we present a wholesale energy market model. Consider a power system network with $M$ buses, $L$ transmission lines, and $G$ generators. Each bus $m \in \{1, \ldots, M\}$ serves $N_m$ household agents, consisting of $N_m^p$ prosumers and $N_m^c$ consumers. Each generator $g \in \{1, \ldots, G\}$ has an associated cost function $C_g(\cdot)$. Since each generator is linked to a specific bus, let $G_m \subseteq \{1, \ldots, G\}$ denote the set of generators located at bus $m$. Clearly, $\cup_{m=1}^M G_m = \{1, \ldots, G\}$ and $G_{m} \cap G_{m'} = \emptyset$ for any distinct $m, m' \in \{1, \ldots, M\}$.

Let $t \in \{1, 2, \ldots\}$ represent the timestep, with each timestep corresponding to a fixed time interval, such as an hour. At each time $t$, an aggregator at bus $m$ collects the bids (which are determined through the MFC approach to be described in the next section) from all the agents under its management and submits the aggregated bids to the system operator. For prosumers, the specific quantity \(d_{imt}\) represents the net demand, defined as the prosumer's demand minus the solar output from their own solar panel outputs. If \(d_{imt} \geq 0\), it indicates net energy demand; if \(d_{imt} < 0\), it represents net supply.

To avoid (unrealistic) scenarios where the net energy supply from prosumers exceeds the total demand of all pure consumers, we assume that the total demand of all agents (both prosumers and consumers) in the system is always greater than or equal to the total PV generation from all prosumers across all $M$ buses. Mathematically, this is expressed as:
\begin{equation}
    \sum_{m=1}^{M} \sum_{i=1}^{N_m} d_{imt} \geq 0, \quad \forall \; t = 1, 2, \ldots
\end{equation}

Upon receiving all bids from power plants, aggregators, and other market participants, the system operator solves an economic dispatch problem to determine the least-cost way of meeting demand from available supply sources while accounting for network constraints. This can be formulated as an optimization problem for each timestep $t$ as follows:
\begin{align}
    \min_{\mathbf{p}_t} \; & \sum_{g=1}^G C_g(p_{gt}) \\
    \text{s.t.} \; & \sum_{g=1}^G p_{gt} = \sum_{m=1}^{M} \sum_{i=1}^{N_m} d_{imt}, \label{eqn:power-balance} \\
    & -\overline{F}_l \leq \sum_{m=1}^{M} PTDF_{lm} \left( \sum_{g \in G_m} p_{gt} - \sum_{i=1}^{N_m} d_{imt} \right) \leq \overline{F}_l, \quad \forall \; l \in \{1, \ldots, L\}, \label{eqn:flow-limit} \\
    & 0 \leq p_{gt} \leq \overline{p}_g, \quad \forall \; g \in \{1, \ldots, G\}, \label{eqn:power-limit}
\end{align}
where $\mathbf{p}_t = \begin{bmatrix} p_{1t} & \cdots & p_{Gt} \end{bmatrix}^T$ represents the vector of power outputs from each generator. The parameter $PTDF_{lm}$ denotes the power transfer distribution factor for line $l$, which connects bus $m$ to a designated hub in a hub-spoke network representation. In this representation, power flow from any bus $m$ to another bus is assumed to flow through the hub, so a line $l$ is considered to connect a bus directly to the hub. $\overline{F}_l$ in \eqref{eqn:flow-limit} is the flow limit for line $l$, and $\overline{p}_g$ in \eqref{eqn:power-limit} is the maximum output capacity of generator $g$. For a fixed network, the parameters $C_g(\cdot)$, $PTDF_{lm}$, $\overline{F}_l$, and $\overline{p}_g$ are known constants. Given the collected bids $d_{imt}$ at each time $t$, the optimization problem can be efficiently solved using an optimization solver.

Upon solving the economic dispatch problem, the LMPs are determined from the dual values associated with the constraints, denoted by $\lambda_{mt}$ for each bus $m$ at time $t$. The dual variables $\lambda^{\text{HUB}}_t, \underline{\mu}_{lt}, \overline{\mu}_{lt}, \underline{\nu}_{gt}, \overline{\nu}_{gt}$ correspond to the Lagrange multipliers for constraints (\ref{eqn:power-balance}), (\ref{eqn:flow-limit}), and (\ref{eqn:power-limit}), respectively. Here, $\lambda^{\text{HUB}}_t$ represents the hub price, while the underlined and overlined $\mu$ and $\nu$ variables represent the dual variables of the lower and upper bounds of the flow and power limits. Defining $D_{mt} = \sum_{i=1}^{N_m} d_{imt}$ as the total demand at bus $m$ at time $t$, the LMP $\lambda_{mt}$ at bus $m$ can be computed as:
\begin{equation}
    \lambda_{mt} \coloneq \frac{\partial \mathcal{L}_t}{\partial D_{mt}} = \lambda^{\text{HUB}}_t - \sum_{l=1}^L PTDF_{lm} (\underline{\mu}_{lt} - \overline{\mu}_{lt}), \quad \forall \; m \in \{1, \ldots, M\},
\end{equation}
where  $\mathcal{L}_t$ is the Lagrangian function of the economic dispatch problem, defined as:
\begin{equation}
    \begin{split}
        \mathcal{L}_t = & \sum_{g=1}^G C_g(p_{gt}) - \lambda^{\text{HUB}}_t \left( \sum_{g=1}^G p_{gt} - \sum_{m=1}^M D_{mt} \right) 
         + \sum_{l=1}^{L} \overline{\mu}_{lt} \left[ \sum_{m=1}^M PTDF_{lm} \left( \sum_{g \in G_m} p_{gt} - D_{mt} \right) - \overline{F}_l \right] \\
        & - \sum_{l=1}^{L} \underline{\mu}_{lt}\left[ \sum_{m=1}^M PTDF_{lm} \left( \sum_{g \in G_m} p_{gt} - D_{mt} \right) + \overline{F}_l \right] 
         + \sum_{g=1}^{G} \left[ \overline{\nu}_{gt} (p_{gt} - \overline{p}_g) - \underline{\nu}_{gt} p_{gt} \right].
    \end{split}
\end{equation}
In essence, the LMP $\lambda_{mt}$ represents the marginal cost of supplying an additional unit of electricity to bus $m$ at time $t$, reflecting both generation costs and network constraints.

\section{Aggregators' Problem -- A Two-Phase Learning Approach}
\label{sec:Agg}
In this section, we consider the problem faced by aggregators managing DERs, specifically prosumers with photovoltaic (PV) panels and energy storage systems. Aggregators develop strategies to optimally charge and discharge these storages under their control using RL algorithms. The RL algorithm employed here is sophisticated, involving a two-phase learning approach. The key elements in presenting the algorithm are listed below.  

\textbf{Time:} At each timestep $t = 0, 1, 2, \ldots$, each aggregator performs reinforcement learning using algorithms such as Proximal Policy Optimization (PPO), Trust Region Policy Optimization (TRPO), Soft Actor-Critic (SAC), and others for $T_{\text{train}}$ steps. We use $\tau = 1, 2, \ldots, T_{\text{train}}$ to denote training timesteps, during which aggregators learn a policy. The actual play phase involves submitting bids and calculating LMPs, and these two phases are detailed in Section \ref{sec:alg}.

\textbf{Prosumers:} We consider a total of $N$ heterogeneous prosumers in this game, where $N = \sum_{m=1}^{M} N^p_m$. Each prosumer $i$ is assigned a fixed type $\theta \in \Theta$ throughout the game, where $\Theta$ is a finite set. Prosumers of the same type $\theta$ share the same net demand distribution, bus location, and reward function. For simplicity, we assume all prosumers at the same bus have the same type, which differs across buses. Thus, the set $\Theta$ corresponds to the set of buses $\{1, \ldots, M\}$.

\textbf{Aggregators:} The assumption of prosumer types allows us to designate an aggregator as a ``representative" for each type. In our setup, each bus has one aggregator who employs an RL algorithm for all prosumers at that bus. Let $\overline{x}_{im}$ denote the storage capacity of the $i$-th prosumer at bus $m$. The aggregator for bus $m$ thus has a total storage capacity of $\overline{x}_m = \sum_{i=1}^{N^p_m} \overline{x}_{im}$.\footnote{For simplicity, we assume that each bus $m$ is served by a single aggregator, which we refer to as Aggregator $m$. This assumption is made for ease of presentation, and relaxing it will not affect any outcomes.} This is the essential idea of the MFC approach as described in \cite{mondal2023mean}, which significantly reduces computational costs since typically $|\Theta|$ is much less than $N$.

\textbf{LMP Beliefs:} Let $H$ be the total number of timesteps in one day. Define two mappings: $h(\cdot)$ and $k(\cdot)$, which map the time index to the corresponding timestep of the day and the day index, respectively. In our case, $h(t) = t \mod H$ and $k(t) = \lfloor t / H \rfloor$. Each aggregator maintains its own belief about the LMPs for each timestep of the day, represented as a vector of length $H$. Let $\hat{\vlambda}_{mt} \in \mathbb{R}^H$ denote the LMP belief of Aggregator $m$ at time $t$. After the system operator solves the economic dispatch problem, the aggregator is updated with the new LMP $\lambda_{mt}$ for its bus and updates its belief according to:
\begin{equation}
    [\hat{\vlambda}_{m,t+1}]_{h(t)} = [\hat{\vlambda}_{mt}]_{h(t)} - \delta \frac{[\hat{\vlambda}_{mt}]_{h(t)} - \lambda_{mt}}{\sqrt{k(t) + 1}},
    \label{eqn:lmp-belief}
\end{equation}
where $\delta \in [0.5, 1]$ is a learning rate hyper-parameter for the LMP update rule.

\textbf{Actions:} Let $\mathcal{A} \subseteq [-1, 1]$ represent the discrete action space. Each action corresponds to a percentage of charging or discharging relative to the storage capacity. At time $\tau$, each aggregator selects an action $a_{\tau} \in \mathcal{A}$. If $a_{\tau} \geq 0$, the aggregator charges the storage by $a_{\tau}$, and if $a_{\tau} < 0$, it discharges $a_{\tau}$. For example, if each action represents 1\%, then $\mathcal{A} = \{-1, -0.99, \ldots, 0, 0.01, \ldots, 0.99, 1\}$.

\textbf{Action Masking:} Action masking is a technique widely used in RL policy gradient algorithms. The idea is to `mask out' invalid actions and sample only from the set of valid actions. This technique has been successfully applied in the context of video games, as demonstrated in \cite{vinyals2017starcraft, berner2019dota, ye2020mastering}, and is theoretically supported in \cite{huang2020closer}. Mathematically, the algorithm assigns a value of $-\infty$ to each invalid action based on the current state, resulting in zero probability for those actions. For instance, if the current storage level is 0.8, any $a < -0.8$ or $a > 0.2$ would be masked out to ensure the storage level remains between 0 and full capacity. From this point onward, all actions are assumed to be valid.

\textbf{States:} Let $\mathcal{S}$ denote the state space. In an RL algorithm, each aggregator takes actions based on their current state. The state at time $\tau$, $s_{\tau} \in \mathcal{S}$, is a tuple consisting of the storage level, net load demand, current LMP belief ($[\hat{\vlambda}_{mt}]_{h(\tau)}$), and the current timestep of the day ($h(\tau)$). The storage level for Aggregator $m$ at time $\tau$ is denoted as $x_{m\tau} \in [0, 1]$. Let $\eta_m \in (0, 1]$ be the storage efficiency parameter. The state transition for Aggregator $m$ after taking action $a_{m\tau}$ is:
\begin{equation}
    x_{m,\tau+1} = x_{m\tau} + \mathbbm{1}_{\{a_{m\tau} < 0\}} \eta_m a_{m\tau} + \mathbbm{1}_{\{a_{m\tau} \geq 0\}} \frac{a_{m\tau}}{\eta_m},
    \label{eqn:storage-transition}
\end{equation}
where $\mathbbm{1}_{\{\cdot\}}$ is the indicator function, equal to 1 if the condition is satisfied and 0 otherwise. To simplify, we define the mapping from an action to the actual percentage of charging or discharging as:
\begin{equation}
    \Phi(a, \eta) = \mathbbm{1}_{\{a < 0\}} \eta a + \mathbbm{1}_{\{a \geq 0\}} \frac{a}{\eta}.
    \label{eqn:act-to-actual}
\end{equation}
Thus, equation (\ref{eqn:storage-transition}) can be rewritten as:
\begin{equation}
    x_{m,\tau+1} = x_{m\tau} + \Phi(a_{m\tau}, \eta_m).
\end{equation}
During training, the aggregator samples the net demand $d_{m\tau}$ (as a percentage relative to the storage capacity) from a load shape profile $Q^p_m$, which is a probability distribution not known to the aggregator, but learned from historical data. 

\textbf{Rewards and Policies:} The single-period reward function is defined as:
\begin{equation}
    r_{\tau} = r(s_{\tau}, a_{\tau}) = - [\hat{\vlambda}_{mt}]_{h(\tau)} \cdot \overline{x}_m \cdot (\Phi(a_{m\tau}, \eta_m) + d_{m\tau}).
\end{equation}
This reward function represents the payoff associated with the aggregator's charging or discharging decisions at each timestep \(\tau\), based on their beliefs of LMPs \([\hat{\vlambda}_{mt}]_{h(\tau)}\), the total storage capacity \(\overline{x}_m\), and the net energy adjustment (as a percentage) \(\Phi(a_{m\tau}, \eta_m) + d_{m\tau}\). 

Through an RL algorithm, each aggregator learns a policy $\pi_{mt}$, a strategy mapping a given state to an action distribution. The objective of RL is to find a policy $\pi_{mt}$ that maximizes the total discounted expected reward:
\begin{equation}
    \pi_{mt} = \arg\max_{\pi_{mt}} \mathbb{E} \left\{ \sum_{\tau=0}^{\infty} \gamma^\tau r_{\tau} \right\},
\end{equation}
where $\gamma \in (0, 1)$ is the discount factor.

\section{The Two-Phase RL Algorithm}
\label{sec:alg}

We now propose a two-phase distributed mean-field RL algorithm based on the setup above. The algorithm comprises two phases: the training phase, where each aggregator trains its policy using RL algorithms for a specified number of steps, and the actual play phase, where all prosumers use their aggregators' trained policies and submit their bids to the system operator to solve an economic dispatch problem and update the LMPs. The algorithm is presented in Algorithm \ref{alg:two-phase}.

\textbf{Training Phase:} At each time $t$, each aggregator fixes its LMP belief and performs RL using an algorithm such as PPO, TRPO, SAC, etc., for $T_{\text{train}}$ steps.

\textbf{Actual Play Phase:} After each aggregator has learned a policy, this policy is distributed to its prosumers. At each bus $m$, each prosumer $i$ samples an action from the policy $a_{imt} \sim \pi_{mt}$. Let $Q^c_m$ denote the net load distribution for consumers at bus $m$, similar to the definition of $Q^p_m$. Each prosumer $i$ and consumer $j$ samples their net load demand from $d_{imt} \sim Q^p_m$ and $d_{jmt} \sim Q^c_m$, respectively. The total demand $D_{mt}$ is then aggregated and submitted to the system operator as a single bid:

\begin{equation}
    D_{mt} = \sum_{i=1}^{N^p_m} (d_{imt} + a_{imt}) + \sum_{j=1}^{N^c_m} d_{jmt}.
    \label{eqn:collect-bids}
\end{equation}

An economic dispatch is then executed to obtain the new LMP $\lambda_{mt}$ for all $m$ at time $t$, which is subsequently used to update the belief as in Equation (\ref{eqn:lmp-belief}).

\textbf{Mean-Field Equilibrium (MFE):} Through mean-field learning, the system can reach an equilibrium point known as an MFE. At an MFE, given a population profile $s^\star$ (mean-field), the corresponding optimal policy is $\pi^\star$; with $\pi^\star$, the population profile remains $s^\star$ after the state transition. Mathematically, an MFE is defined as follows:

\begin{definition}
    An optimal policy $\pi^\star$ and a population profile $s^\star$ form an MFE if 
    \begin{equation}
        s^\star(\cdot) = \int_{\mathcal{S}} \int_{\mathcal{A}} \mathbb{P}(\cdot | s^\star, a) \, d\pi^\star(s^\star, s)(a) \, ds^\star(s),
    \end{equation}
\end{definition}
where \(\mathbb{P}(\cdot | s^\star, a)\) represents the state transition probability, which is the probability distribution over the next state, given the current state \(s^\star\) and action \(a\) and defines the dynamics of the system within the mean-field framework, as it determines how the population profile changes in response to the strategies of the agents.

In our multi-agent RL problem, the information about the mean-field is embedded in the LMPs through economic dispatch. Thus, we can treat the LMP $\vlambda$ as a function of the mean-field and claim that an MFE is achieved when a stationary pair of $\pi^\star$ and $\vlambda^\star$ emerge. In Section \ref{sec:numerical}, we empirically demonstrate that the MFE can be reached.

\begin{algorithm}
    \caption{A two-phase distributed mean-field RL algorithm for prosumers to bid energy and update the belief of LMP}
    \label{alg:two-phase}
    \KwIn{Initial battery state $x_{m0} \in [0, 1]$; initial LMP belief $\hat{\vlambda}_{m0} \in \mathbb{R}^H$; LMP belief learning parameters $\delta_m$; net demand shapes for prosumers and consumers $Q^p_m, Q^c_m$ for each bus $m = 1, \ldots, M$; training steps $T_{\text{train}}$.}
    \For{$t = 0, 1, \ldots$}{
        \Comment{Initialization}
        Get the timestep of the day $ h \gets t \mod H $ \;
        Get the day index $ k \gets \lfloor t / H \rfloor $\;
        Initialize an empty bid collector $ D_t \gets \{ \} $ \;

        \Comment{Training phase}
        \ForEach{bus $m = 1, \ldots, M$}{
            Use RL to train the aggregator for $T_{\text{train}}$ steps with initial storage $x_{mt}$ under $\hat{\vlambda}_{mt}$ and obtain policy $\pi_{mt}$ \;
        }

        \Comment{Actual play phase}
        \ForEach{bus $m = 1, \ldots, M$}{
            $D_{mt} \gets 0$ \;
            \ForEach{Prosumer $i = 1, \ldots, N^p_m$}{
                Sample net demand $d_{imt} \sim Q^p_m$ \;
                Sample charging/discharging actions $a_{imt} \sim \pi_{mt}$ \;
            }
            \ForEach{Consumer $j = 1, \ldots, N^c_m$}{
                Sample net demand $d_{jmt} \sim Q^c_m$ \;
            }
            Collect all bids $D_{mt} \gets \sum_{i=1}^{N^p_m} (d_{imt} + \Phi(a_{imt}, \eta_{im})) + \sum_{j=1}^{N^c_m} d_{jmt}$ as in (\ref{eqn:collect-bids}) \;
            Submit the bid to collector $D_t \gets D_t \cup \{ D_{mt} \overline{x}_{m} \}$ \;
            Storage state transition: $x_{m,t+1} \gets x_{mt} + \frac{1}{N^p_m} \sum_{i=1}^{N^p_m} \Phi(a_{imt}, \eta_{im})$ \;
        }
        Solve economic dispatch and get $\lambda_{mt}$ \;
        \ForEach{bus $m = 1, \ldots, M$}{
            Make a copy of the LMP belief: $\hat{\vlambda}_{m,t+1} \gets \hat{\vlambda}_{mt}$ \;
            Update the $h$-th belief only: $[\hat{\vlambda}_{m,t+1}]_h \gets [\hat{\vlambda}_{mt}]_h - \delta_m \frac{[\hat{\vlambda}_{mt}]_h - \lambda_{mt}}{\sqrt{k + 1}} $ as in (\ref{eqn:lmp-belief}) \;
        }
    }
\end{algorithm}

\section{Numerical Experiment} 
\label{sec:numerical}

In this section, we conduct numerical experiments to analyze market behavior under the mean-field game framework described earlier.

\subsection{Test Case}

We utilize the 37-bus synthetic network from \cite{birchfield}, which corresponds to the geographical layout of the Hawaiian island of Oahu. We modify this case by mapping each power plant from Hawaiian Electric \cite{powerfacts} to its nearest bus. After modification, the network comprises a total of 26 generators: 4 oil, 2 biomass, 17 (grid-scale) solar, and 3 wind generators. We assume quadratic cost functions for all oil and biomass generators, applying coefficients from \cite{isone, tidball2010cost} to these respective generator types. The ranges for these coefficients are presented in Table \ref{tab:cost-coeff}. Additionally, the generation cost for solar and wind generators is set to zero, with actual outputs adjusted by a capacity factor based on weather conditions. Let $\Delta(a,b,c)$ denote a triangular distribution with lower limit $a$, upper limit $b$, and mode $c$. In our simulation, we assume that solar generation follows an average profile from \cite{profilesolar}, scaled by a factor following $\Delta(0.8, 1.2, 1)$, while wind generation is scaled by a factor following $\Delta(0.5, 1.5, 1)$ from \cite{argueso}.

\begin{table}
    \caption{Cost function coefficient ranges for non-renewable fuel types ($C(p) = ap^2 + bp$)}
    \label{tab:cost-coeff}
    \begin{center}
        \begin{tabular}{c | c c} 
            \hline
            Fuel Type & $a \, (\$/\text{MW}^2\text{h})$  & $b \, (\$/\text{MWh})$ \\ 
            \hline
            Oil & 0.0059 - 0.0342 & 19.98 \\ 
            Biomass & 0.001 - 0.002 & 28.45 - 52.65 \\
            \hline
        \end{tabular}
    \end{center}
\end{table}

\begin{figure}[ht]
    \centering
    \includegraphics[width=0.5\linewidth]{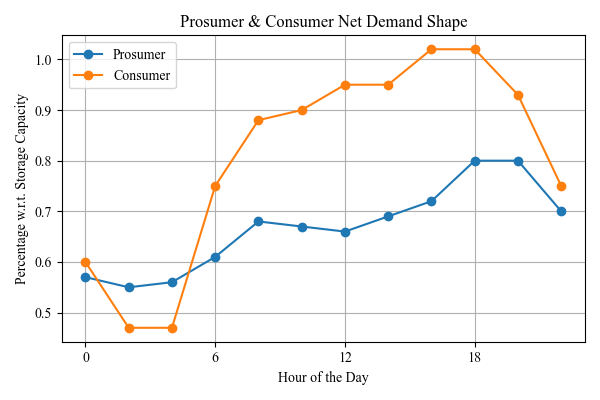}  
    \caption{Net demand shape for prosumers and consumers as a percentage with respect to energy storage capacity (10 kWh). Data is adapted from \cite{coffman2016estimating}.}
    \label{fig:load-shape}
\end{figure}

Each bus in the network hosts two types of agents: prosumers with DERs and energy storage, and pure consumers. Each energy storage unit is set to a capacity of 10 kWh. The daily demand profiles for prosumers and consumers are derived from average hourly net demand (gross load minus DER generation) and gross load data from \cite{coffman2016estimating}, respectively. We further assume that actual demand is the average demand scaled by a factor following $\Delta(0.9,1.1,1)$. Figure \ref{fig:load-shape} illustrates the net demand shape for prosumers and consumers at each timestep of the day.

\subsection{Results}

\begin{figure}[ht]
    \begin{center}
        \begin{subfigure}{.49\textwidth}
          \centering
          \includegraphics[width=\linewidth]{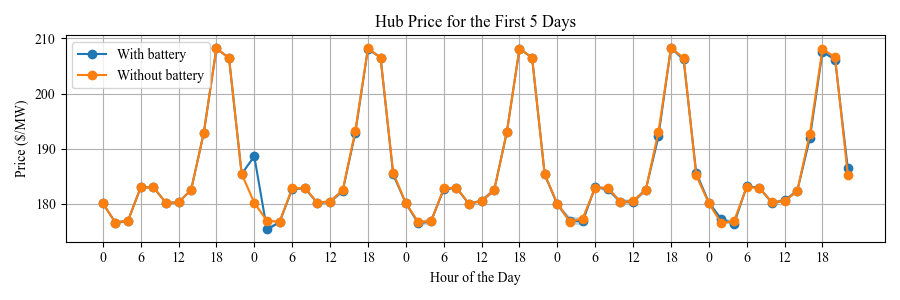}  
          \caption{Hub prices over the first 5 days during learning}
          \label{fig:hub-prices-first}
        \end{subfigure}
        \begin{subfigure}{.49\textwidth}
          \centering
          \includegraphics[width=\linewidth]{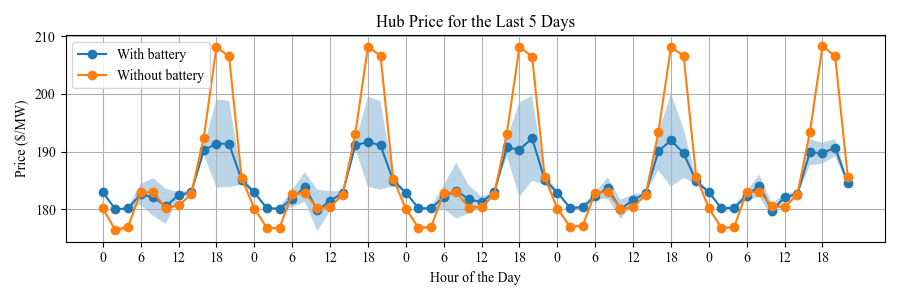}  
          \caption{Hub prices over the last 5 days during learning}
          \label{fig:hub-prices-last}
        \end{subfigure}
        \caption{Comparison of hub prices with and without energy storage}
        \label{fig:hub-prices}
    \end{center}
\end{figure}

\begin{figure}[ht]
    \begin{center}
        \begin{subfigure}{.49\textwidth}
          \centering
          \includegraphics[width=\linewidth]{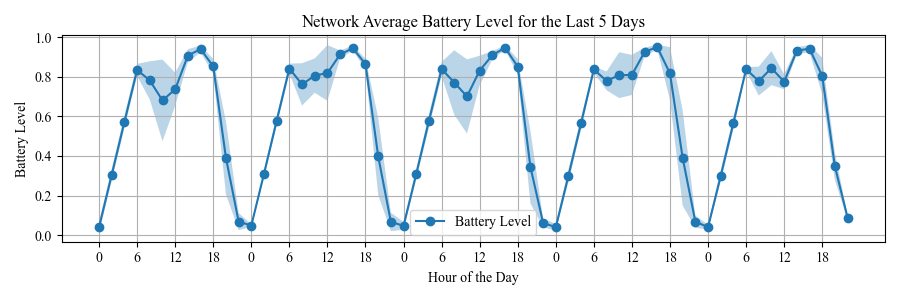}  
          \caption{Average storage level across the network}
          \label{fig:storages-states}
        \end{subfigure}
        \begin{subfigure}{.49\textwidth}
          \centering
          \includegraphics[width=\linewidth]{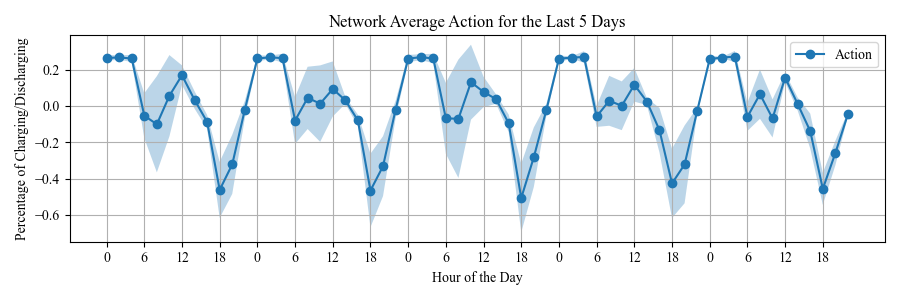}  
          \caption{Average charging/discharging actions across the network}
          \label{fig:storages-actions}
        \end{subfigure}
        \caption{Average storage levels and charging/discharging actions over the last 5 days across the network}
        \label{fig:storages}
    \end{center}
\end{figure}

\begin{figure}[ht]
    \begin{center}
        \begin{subfigure}{.49\textwidth}
          \centering
          \includegraphics[width=\linewidth]{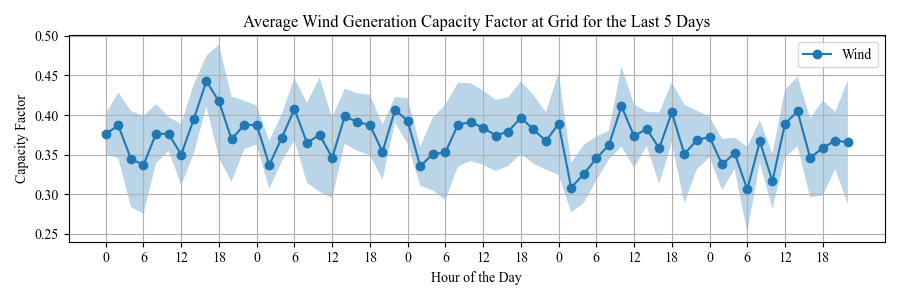}  
          \caption{Wind generator capacity factor}
          \label{fig:renewables-wind}
        \end{subfigure}
        \begin{subfigure}{.49\textwidth}
          \centering
          \includegraphics[width=\linewidth]{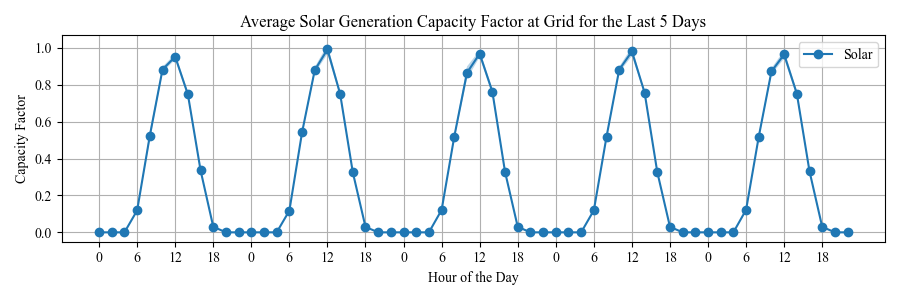}  
          \caption{Solar generator capacity factor}
          \label{fig:renewables-solar}
        \end{subfigure}
        \caption{Capacity factors for renewable generators (wind and solar) over the last 3 days. Wind data is adapted from \cite{argueso}, and solar data from \cite{profilesolar}.}
        \label{fig:renewables}
    \end{center}
\end{figure}

We employ PPO as the learning algorithm and set $T_{\text{train}} = 3,600$, with $H=12$ (corresponding to 2-hour timesteps). The simulation is run for a period of 50 days and repeated 5 times using different random seeds. The experiments were conducted on a Windows 11 system equipped with a 13th Gen Intel(R) Core(TM) i7-13700KF (24 cores) and NVIDIA GeForce RTX 4070. Figures \ref{fig:hub-prices}, \ref{fig:storages}, and \ref{fig:renewables} display the hub prices, storage levels and charging/discharging actions, and renewable generators' capacity factors, respectively.

Additionally, to study the impacts of energy storage coupled with the RL algorithms on LMP volatility, we adopt the incremental mean volatility (IMV) measure from \cite{roozbehani} as the metric. We also compare the average cumulative \emph{ex-post} cost (the product of the bidding quantity and the real-time LMP) across all buses. The IMV of a sequence of LMPs $\{\lambda_t\}_{t=1}^{\infty}$ is defined as:
\begin{equation}
  IMV = \lim_{T \to \infty} \frac{1}{T} \sum_{t=1}^T \abs{\lambda_{t+1} - \lambda_{t}}.
  \label{eqn:measure}
\end{equation}

\begin{figure}[ht]
    \begin{center}
        \begin{subfigure}{.49\textwidth}
          \centering
          \includegraphics[width=\linewidth]{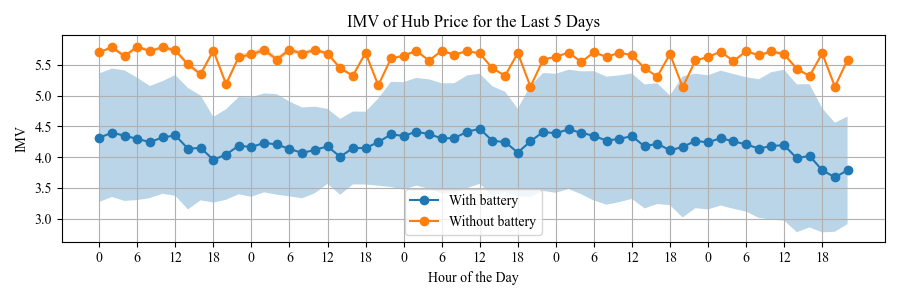}  
          \caption{IMV of the hub price over 5 runs during the last 5 days}
          \label{fig:measure-imv}
        \end{subfigure}
        \begin{subfigure}{.49\textwidth}
          \centering
          \includegraphics[width=\linewidth]{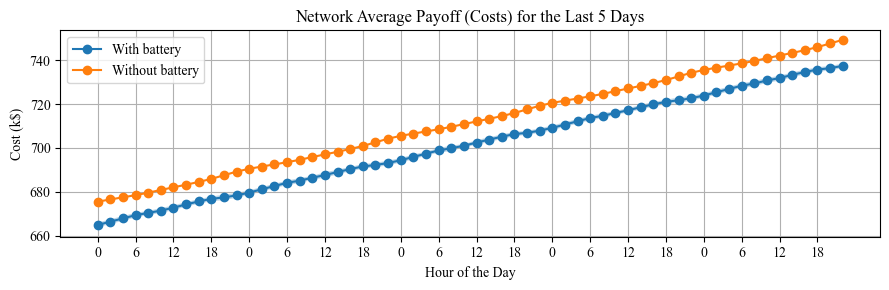}  
          \caption{Cumulative \emph{ex-post} costs over the last 5 days}
          \label{fig:measure-cost}
        \end{subfigure}
        \caption{Comparison of IMV and cumulative \emph{ex-post} costs between learning (with storage) and non-learning (without storage) over 5 runs during the last 5 days}
        \label{fig:measure}
    \end{center}
\end{figure}

We approximate the IMV for a sequence of LMPs and calculate the cumulative \emph{ex-post} cost over 5 simulation runs. Figure \ref{fig:measure} shows the results for IMVs and costs over the last 5 days. It is evident that learning reduces both volatility and cumulative \emph{ex-post} costs.

\subsection{Discussion}

Our results indicate that prosumers can learn strategies to charge when prices are low and discharge when prices are high using our algorithm. As shown in Figure \ref{fig:hub-prices-last}, hub prices stabilize towards an equilibrium, and prosumer actions illustrated in Figure \ref{fig:storages-actions} align with a mean-field equilibrium (MFE). Some fluctuations are observed at the 9th timestep in a day, likely due to variations in the solar/wind capacity factor. Since prices are determined by the next cheapest generator available to supply an additional unit of power, fluctuations in renewable capacity factors can lead to increased or decreased reliance on non-renewable sources, making prices sensitive to small changes in charging/discharging actions. Overall, our model demonstrates the potential for mitigating extreme price swings and fostering a more stable market environment under decentralized decision-making.

\section{Conclusion}\label{sec:Conclusion}

In this paper, we propose a decentralized, mean-field-based model where aggregators make informed charging and discharging decisions on behalf of the prosumers they manage. These decisions are guided by LMPs in a wholesale energy market, which encapsulate information about aggregate demand, supply dynamics, and network constraints. Additionally, we introduce a two-phase learning algorithm within the hybrid MFC and MFG framework, leveraging RL to iteratively learn optimal policies for aggregators while facing various uncertainties from renewable output, demand fluctuations, and evolving market dynamics. 

Our numerical experiments suggest that the proposed approach can achieve convergence to an MFE, where both the aggregators' strategies and the LMPs stabilize. Additionally, comparisons between scenarios with and without energy storage show that integrating energy storage with our RL-based algorithms can effectively reduce extreme LMP volatility, promoting a more stable market environment under decentralized, aggregator-led decision-making. This approach also provides cost benefits to consumers by optimizing the use of energy storage.

Future work will focus on providing a theoretical foundation for our model, specifically proving the convergence of the two-phase RL algorithm to an MFE. Additionally, we plan to explore the use of different RL algorithms to further enhance performance under various market conditions.

\printbibliography

\begin{lstlisting}
	
\end{lstlisting}

\end{document}